\documentstyle[preprint,prd,aps]{revtex}
\begin{document}
\draft
\preprint{}
\newcommand{\be}{\begin{equation}}
\newcommand{\ee}{\end{equation}}
\newcommand{\bea}{\begin{eqnarray}}
\newcommand{\eea}{\end{eqnarray}}
\title{Gauge Independent Trace Anomaly for Gravitons}
\author{
H. T. Cho \footnote[1]{bsp15@twntku10.tku.edu.tw }}
\address{Tamkang University, Department of Physics,\\
Tamsui, Taipei, TAIWAN 251, R.O.C.}
\vskip -.5 truein
\author{
R. Kantowski \footnote[2]{kantowski@phyast.nhn.uoknor.edu}}
\address{
University of Oklahoma, Department of Physics and Astronomy,\\
Norman, OK 73019, USA
}
\maketitle
\begin{abstract}
We show that the trace anomaly for gravitons
calculated using the
usual effective action formalism depends on the choice
of gauge when the background spacetime is not a solution
of the classical equation of motion, that is, when
off-shell. We then use the gauge independent
Vilkovisky-DeWitt effective action to restore gauge independence
to the off-shell case.
Additionally we explicitly evaluate trace anomalies
for some N-sphere background spacetimes.
\end{abstract}

\section{Introduction}
The usual effective action, in general, depends on the choice
of quantum gauge
fixing when the background field is not a solution of the
classical equations
of motion, that is, when the background is off-shell \cite{GK}.
This has posed a problem
in the use of the effective action formalism to study, for example,
the spontaneous compactification of Kaluza-Klein spaces \cite{R-DS,KL} (see
\cite{BOS} for additional references).
The same gauge dependence
problem exists in the calculation of the trace anomaly for
gravitons \cite{RC}
because both the background spacetime and the graviton
fields stem from the metric. When the background
spacetime is not a solution of the Einstein equations, the
trace anomaly calculated from the usual effective action may
depend on the gauge choice.
We illustrate this explicitly in Section II
for the simple case of Einstein gravity with cosmological
constant in a flat background.

To overcome the gauge dependency we use the Vilkovisky-DeWitt
(VD) effective
action formalism \cite{GV,BSD}. The VD formalism has
been applied
to spontaneous compactification of Kaluza-Klein spaces
and unique answers which are independent of the
choice of gauge have been obtained \cite{BO,HKLT,BLO,CK1}.
Recently, it has also been used to study 2-d quantum gravity
\cite{KM} and even gravity-GUT unifications \cite{O}. In Section III we define
the unique
trace anomaly for gravitons using the one-loop VD effective
action.  We evaluate this VD trace anomaly for
the case considered in Section II and show that it is indeed
independent of gauge choice.

For most gauges the VD effective action involves evaluating
the determinants or the $\zeta$-functions
of complicated non-local operators. However, the calculation
simplifies when the Landau-DeWitt gauge is used. Since the
VD effective action is independent of gauge choice, one can of
course choose whatever gauge convenient without altering the final
results. In this particular gauge the operators become local,
but remain non-minimal \cite{BV}. In a previous paper \cite{CK}
we have devised
a method to evaluate the $\zeta$-functions (at argument 0) of
non-minimal vector and
tensor operators on maximally symmetric spaces. In Section IV
we use this method  to calculate the VD trace anomalies
for such background
spaces. Explicit results for N-spheres and Euclidean spaces
of dimensions 4, 6, 8, and 10 are given. Finally, conclusions
are given in Section V.

\section{Gauge Dependence of The Trace Anomaly}

In this section we demonstrate the dependence  of the trace
anomaly for gravitons on the quantum gauge
choice. To do so we
consider the simple case of Einstein gravity with a
cosmological constant in a flat background spacetime.
The corresponding action is (in Euclidean signature),
\[
S\equiv \int d^4x{\cal{L}},
\]
where
\begin{equation}
{\cal{L}}=-\sqrt{\bar{g}}(\bar{R}-2\Lambda)\ ,
\label{action}
\end{equation}
(see Eqs. (\ref{Riemann})-(\ref{RicciS}) for curvature conventions).
The metric is split into its background and quantum
parts:
\begin{equation}
\bar{g}_{\mu\nu}=\delta_{\mu\nu}+h_{\mu\nu}\ ,
\label{split}
\end{equation}
where $h_{\mu\nu}$ is the graviton field. To evaluate the trace
anomaly for gravitons, we expand the Lagrangian in powers of
$h_{\mu\nu}$ keeping only the quadratic part,
\begin{equation}
{\cal{L}}_{quad}=\frac{1}{4}h_{\mu\nu ,\rho}h_{\mu\nu ,\rho}
-\frac{1}{8}h_{,\rho}h_{,\rho}
-\frac{1}{2}(h_{\mu\rho ,\mu}-\frac{1}{2}h_{,\rho})
(h_{\rho\nu ,\nu}-\frac{1}{2}h_{,\rho})
-\frac{1}{2}(h_{\mu\nu}^{2}-\frac{1}{2}h^{2})\Lambda\ ,
\label{lquad}
\end{equation}
where $h\equiv h_{\mu\mu}$. Next, we introduce the
gauge-fixing term and the corresponding ghost term.
We choose
a one-parameter ($\alpha$) family of covariant gauges,
\begin{equation}
{\cal{L}}_{gf}=\frac{1}{2\alpha}(h_{\mu\rho ,\mu}
-\frac{1}{2}h_{,\rho})(h_{\rho\nu ,\nu}
-\frac{1}{2}h_{,\rho})\ .
\label{lgf}
\end{equation}

The corresponding ghost Lagrangian is
\begin{equation}
{\cal{L}}_{gh}=\overline{\eta}_{\mu}(-\partial^{2})\eta_{\mu}\ ,
\label{lgh}
\end{equation}
where $\overline{\eta}_{\mu}$ and $\eta_{\mu}$ are vector ghosts.
Putting the Lagrangians Eqs.~(\ref{lquad})-(\ref{lgh})
together, we have the quantum Lagrangian for gravitons,
\begin{eqnarray}
{\cal{L}}_{q}&=&\frac{1}{4}h_{\mu\nu}(-\partial^{2}-2\Lambda)
h_{\mu\nu}-\frac{1}{4}h\left[\left(1-\frac{1}{2\alpha}\right)
(-\partial^{2})-\Lambda\right]h
\nonumber\\
&&\ \ +\frac{1}{2}h_{\mu\nu}\left(1-\frac{1}{\alpha}\right)
\partial_{\nu}\partial_{\rho}h_{\mu\rho}
-\frac{1}{2}h\left(1-\frac{1}{\alpha}\right)
\partial_{\mu}\partial_{\nu}h_{\mu\nu}
+\overline{\eta}_{\mu}(-\partial^{2})\eta_{\mu}\ .
\label{lq}
\end{eqnarray}

We use $\zeta$-function regularization to evaluate
the trace anomaly. The $\zeta$-function of
an operator $M$ is defined by
\begin{equation}
\zeta_{M}(s)\equiv\sum_{\lambda}\lambda^{-s}\ ,
\end{equation}
where $\lambda$'s are the eigenvalues of the operator $M$.
Thus we first have to find the eigenvalues of the operators acting
on $h_{\mu\nu}$ and on the ghost fields from Eq.~(\ref{lq}). To
do so we rewrite this Lagrangian as,
\begin{equation}
{\cal{L}}_{q}=\frac{1}{2}\psi_{i}\Theta_{ij}\psi_{j}
+\overline{\eta}^{\mu}(-\partial^{2})\eta_{\mu}\ ,
\end{equation}
where $\psi_{i}$, $i=1,...,10$, are the ten independent
components of $h_{\mu\nu}$. The eigenvalues of the
matrix $\Theta_{ij}$ are \cite{HTC1},
\begin{eqnarray}
&&\lambda_{1}=\lambda_{2}=\lambda_{3}=k^{2}-2\Lambda\ ,
\label{lambda1}\\
&&\lambda_{4}=\lambda_{5}=\frac{1}{2}(k^{2}-2\Lambda)\ ,
\label{lambda4}\\
&&\lambda_{6}=\lambda_{7}=\lambda_{8}=\frac{1}{\alpha}
(k^{2}-2\alpha\Lambda)\ ,
\label{lambda6}\\
&&\lambda_{9}=\frac{1}{2\alpha}\left\{(1-\alpha)k^{2}
+\sqrt{(1-\alpha+\alpha^{2})k^{4}-2\alpha\Lambda(1+\alpha)
k^{2}+4\alpha^{2}\Lambda^{2}}\right\}\ ,
\label{lambda9}\\
&&\lambda_{10}=\frac{1}{2\alpha}\left\{(1-\alpha)k^{2}
-\sqrt{(1-\alpha+\alpha^{2})k^{4}-2\alpha\Lambda(1+\alpha)
k^{2}+4\alpha^{2}\Lambda^{2}}\right\}\ ,
\label{lambda10}
\end{eqnarray}
having been written in momentum space.
Hence, the $\zeta$-function for the graviton field $h_{\mu\nu}$
is,
\begin{equation}
\zeta_{gr}(s)=\sum^{10}_{i=1}\sum_{k}\lambda^{-s}_{i}\ ,
\label{zetagr}
\end{equation}
and the $\zeta$-function for the ghost fields is,
\begin{equation}
\zeta_{gh}(s)=\sum_{k}(k^{2})^{-s}\ .
\label{zetagh}
\end{equation}

Following the arguments similar to \cite{RC} and
\cite{LP}, for example,
it is easy to see that the trace anomaly for gravitons
is given by,
\begin{equation}
\langle{T_{\mu}}^{\mu}\rangle=\frac{1}{V_{4}}\left[\zeta_{gr}(0)
-2\zeta_{gh}(0)\right]\ ,
\label{trace}
\end{equation}
where $V_{4}$ is the volume of the spacetime. $\zeta(0)$ can be
calculated as follows. For a general eigenvalue $ak^{2}+b$,
where $a$ and $b$ are constants,
\begin{eqnarray}
\zeta(0)&=&\lim_{s\rightarrow 0}\sum_{k}(ak^{2}+b)^{-s}\ ,
\nonumber\\
&=&\lim_{s\rightarrow 0}V_{4}\int\frac{d^{4}\!k}{(2\pi)^{4}}
\frac{1}{\Gamma(s)}\int^{\infty}_{0}d\tau\ \tau^{s-1}
e^{-\tau(ak^{2}+b)}\ .
\label{zeta}
\end{eqnarray}
Evaluating the $k$-integral first and then the $\tau$-integral,
\begin{eqnarray}
\zeta(0)&=&\lim_{s\rightarrow 0}
\frac{V_{4}}{\Gamma(s)}\int^{\infty}_{0}d\tau\
\tau^{s-1}e^{-\tau b}\left(\frac{1}{4\pi a\tau}\right)^{2}\ ,
\nonumber\\
&=&\lim_{s\rightarrow 0}\frac{V_{4}}{(4\pi)^{2}}
\frac{\Gamma(s-2)}{\Gamma(s)}\frac{b^{2-s}}{a^{2}}\ ,
\nonumber\\
&=&\frac{V_{4}}{(4\pi)^{2}}\left(\frac{1}{2}\right)
\left(\frac{b}{a}\right)^{2}\ .
\end{eqnarray}
Using this result, we see that for $\lambda_{1}$ through $\lambda_{8}$
in Eqs.~(\ref{lambda1}) through (\ref{lambda6}),
\begin{eqnarray}
\lim_{s\rightarrow 0}\sum_{i=1}^{8}\sum_{k}\lambda_{i}^{-s}
&=&\frac{V_{4}}{(4\pi)^{2}}\left(\frac{1}{2}\right)
\left[3(-2\Lambda)^2+2(-2\Lambda)^{2}+3(-2\alpha\Lambda)^{2}\right]\ ,
\nonumber\\
&=&\frac{V_{4}}{(4\pi)^2}(5+3\alpha^{2})(2\Lambda^2)\ ,
\label{zetagr1}
\end{eqnarray}
and for Eq.~(\ref{zetagh}),
\begin{equation}
\zeta_{gh}(0)=0\ .
\label{zetaghf}
\end{equation}
For $\lambda_{9}$ and $\lambda_{10}$ it is more difficult
to evaluate the $k$-integral because the eigenvalues are not
polynomials in
$k^{2}$. However, we are only interested in the $\zeta$-functions
at $s=0$, and this depends only on the small $\tau$ behavior in
the integrand, behavior which is much like the $\tau$-integral
in Eq.~(\ref{zeta}).
To extract the small $\tau$ behavior from the integral over
$k$, we need only concentrate on the large $k$ behavior. Hence,
we can expand $\lambda_{9}$ and $\lambda_{10}$ as power series
in $1/k^{2}$, and then evaluate the integrals to obtain $\zeta(0)$.
This is done in detail in \cite{CK}. Following the procedures there,
we have
\begin{equation}
\lim_{s\rightarrow 0}\sum_{k}\left(\lambda_{9}^{-s}
+\lambda_{10}^{-s}\right)=
\frac{V_{4}}{(4\pi)^2}(1+\alpha^{2})(2\Lambda^{2})\ .
\label{zetagr2}
\end{equation}
Combining the results in Eqs.~(\ref{zetagr1}) and (\ref{zetagr2}),
we have
\begin{equation}
\zeta_{gr}(0)=\frac{V_{4}}{(4\pi)^2}(3+2\alpha^{2})(4\Lambda^{2})\ .
\label{zetaf}
\end{equation}
Then from Eq.~(\ref{trace}) the trace anomaly for gravitons in
a flat background spacetime with the one-parameter covariant gauge
of Eq.~(\ref{lgf}) is
\begin{equation}
\langle{T_{\mu}}^{\mu}\rangle=\frac{1}{(4\pi)^{2}}(3+2\alpha^{2})
(4\Lambda^{2})\ .
\label{tracef}
\end{equation}
We see that this trace anomaly depends on the gauge parameter
$\alpha$. For example, in the Landau-DeWitt gauge,
$\alpha=0$,
\begin{equation}
\langle{T_{\mu}}^{\mu}\rangle=\frac{1}{(4\pi)^{2}}(12\Lambda^{2})\ ,
\end{equation}
and in the Feynman gauge, $\alpha=1$,
\begin{equation}
\langle{T_{\mu}}^{\mu}\rangle=\frac{1}{(4\pi)^{2}}(20\Lambda^{2})\ .
\end{equation}
This happens because the flat background is not a classical
solution of Einstein gravity with a cosmological constant. The
usual effective action for an off-shell background is in general
dependent on what quantum gauge fixing is used. It is therefore,
not surprising to see that
the trace anomaly calculated from this effective action
also depends on the choice of gauge fixing. In the next section
we remedy this
by
adopting the gauge-independent VD effective action formalism,
and define the unique trace anomaly for the
off-shell case.

\section{Gauge Independence of the Vilkovisky-DeWitt Trace Anomaly}

In this section we introduce necessary elements of the VD effective
action formalism. We
then calculate the trace anomaly for the flat space case
considered in the previous section and show
that it is indeed independent of gauge choice.

To establish notation we write the conventional one-loop effective
action as
\begin{equation}
\Gamma_{1}[\phi]=S[\phi]-\frac{1}{2}{\rm Tr\ ln}S_{,ij}[\phi]\ ,
\end{equation}
where $\phi$ is now a general background field which may not be
a solution of the classical equation of motion and $S$ is defined in
Eq. (\ref{action}). Note that we have
used a condensed notation where $i$ represents both discrete and
continuous indices and $S_{,i}$ denotes a functional derivative.
The VD effective action can be obtained simply be replacing the
ordinary derivative with the covariant functional derivative:
\begin{equation}
S_{,ij}\rightarrow S_{;ij}=S_{,ij}-\Gamma^{k}_{ij}S_{,k}\ ,
\end{equation}
where $\Gamma^{k}_{ij}$ is the connection of the field space. For
nongauge theories, the connection can be constructed from the
metric $G_{ij}$ of this field space. It is just the Christoffel
connection
\begin{equation}
\left\{
\begin{array}{c}
k \\ i\ j
\end{array}
\right\}=\frac{1}{2}G^{kl}(G_{li,j}+G_{lj,i}-G_{ij,l})\ .
\label{christ}
\end{equation}
The prescription for defining $G_{ij}$ has been given by
Vilkovisky \cite{GV}.

For gauge theories, the connection on the
physical field space is the Christoffel connection modulo local gauge
transformations. Let $Q^{i}_{\alpha}$ be the generator of the
gauge symmetry:
\begin{equation}
\delta\phi^{i}=Q^{i}_{\alpha}\epsilon^{\alpha}\ ,
\end{equation}
where $\epsilon^{\alpha}$ are parameters for the transformations.
Then
\begin{equation}
\gamma_{\alpha\beta}=G_{ij}Q^{i}_{\alpha}Q^{j}_{\beta}
\label{gmetric}
\end{equation}
is the metric on that part of the field space
orthogonal to the physical directions. The connection
$\Gamma^{k}_{ij}$ for the physical field space is given
by
\begin{equation}
\Gamma^{k}_{ij}=\left\{
\begin{array}{c}
k \\ i\ j
\end{array}
\right\}+T^{k}_{ij}\ ,
\label{gamma}
\end{equation}
where
\begin{eqnarray}
T_{ij}^{k}&=&-2Q^{k}_{\alpha;(i}Q^{\alpha}_{j)}+
Q^{l}_{\sigma}Q^{k}_{\rho;l}Q^{\sigma}_{(i}Q^{\rho}_{j)}\ ,
\label{tconnect}
\\
Q^{\alpha}_{i}&=&G_{ij}\gamma^{\alpha\beta}Q^{j}_{\beta}\ .
\label{qgauge}
\end{eqnarray}
The derivation of Eq.~(\ref{gamma}) can be found in \cite{GK}. Note
that there is a factor of $\frac{1}{2}$ in the symmetrization.
The covariant derivatives in Eq.~(\ref{tconnect}) are with respect
to the Christoffel connection.
The VD one-loop effective action is now given by
\begin{equation}
\Gamma_{VD}[\phi]=S[\phi]-\frac{1}{2}{\rm Tr\ ln}
\left[G^{li}\left(S_{,ij}-\left\{
\begin{array}{c}
k \\ i\ j
\end{array}
\right\}S_{,k}-T^{k}_{ij}S_{,k}\right)\right]\ ,
\end{equation}
plus contributions from the ghost determinant. In this definition
$\Gamma_{VD}$ is a scalar on the physical field space.
A change of gauge conditions corresponds to just
a coordinate transformation of the physical field space and
leaves $\Gamma_{VD}$ invariant.

We now return
to the problem in Section II of calculating the
trace anomaly for gravitons in a flat background spacetime.
First we evaluate the Christoffel symbols and
$T_{ij}^{k}$ in Eq.~(\ref{gamma}). Following Vilkovisky
\cite{GV}, we take the metric for the field space of
metrics as,
\begin{equation}
G_{g_{\mu\nu}(x)g_{\alpha\beta}(y)}=\frac{1}{4}
\sqrt{g}(g^{\mu\alpha}g^{\nu\beta}+g^{\mu\beta}g^{\nu\alpha}
-g^{\mu\nu}g^{\alpha\beta})\delta^{4}(x-y)\ .
\label{fmetric}
\end{equation}
The Christoffel symbol (see Eq.~(\ref{christ})) is thus \cite{HTC1},
\begin{eqnarray}
&&\left.\left\{
\begin{array}{c}
g_{\rho\sigma}(z) \\ g_{\mu\nu}(x)g_{\alpha\beta}(y)
\end{array}
\right\}\right\vert_{back}
\nonumber\\
&=&-\frac{1}{2}
\delta^{4}(x-z)\delta^{4}(y-z)
\biggl[\delta_{\alpha(\mu}\delta_{\nu)(\rho}\delta_{\sigma)\beta}
+\delta_{\beta(\mu}\delta_{\nu)(\rho}\delta_{\sigma)\alpha}
-\frac{1}{2}\delta_{\alpha\beta}\delta_{\mu(\rho}
\delta_{\sigma)\nu}
\nonumber\\
&&\ \ \ \ -\frac{1}{2}
\delta_{\mu\nu}\delta_{\alpha(\rho}\delta_{\sigma)\beta}
-\frac{1}{4}\delta_{\rho\sigma}
(\delta_{\mu\alpha}\delta_{\nu\beta}+\delta_{\mu\beta}
\delta_{\nu\alpha}-\delta_{\mu\nu}\delta_{\alpha\beta})\biggr]\ ,
\label{christb}
\end{eqnarray}
where $\vert_{back}$ means that the quantity is evaluated at the
background value. Now from Eq.~(\ref{action}),
\begin{equation}
\left.S_{,g_{\mu\nu}(x)}\right\vert_{back}=\Lambda\delta_{\mu\nu}\ ,
\end{equation}
and combining with Eq.~(\ref{christb}), we have
\begin{equation}
\left.{\left\{
\begin{array}{c}
g_{\rho\sigma}(z) \\ g_{\mu\nu}(x)g_{\alpha\beta}(y)
\end{array}
\right\}S_{,g_{\rho\sigma}(z)}}\right\vert_{back}=0\ .
\label{christt}
\end{equation}
Therefore in the case of a flat background the only possible VD
correction comes from the
$T_{ij}^{k}$ term.

To evaluate $T_{ij}^{k}$, we need to know the generators of
the gauge symmetry. For metric fields  gauge symmetry is general
coordinate covariance:
\begin{equation}
\delta g_{\mu\nu}=-g_{\mu\alpha}\partial_{\nu}\epsilon^{\alpha}
-g_{\alpha\nu}\partial_{\mu}\epsilon^{\alpha}
-\epsilon^{\alpha}\partial_{\alpha}g_{\mu\nu}\ ,
\end{equation}
for some set of gauge parameters $\epsilon^{\alpha}$. The generators are thus,
\begin{equation}
Q^{g_{\mu\nu}(x)}_{\alpha y}=-g_{\mu\alpha}\partial_{\nu}
\delta^{4}(x-y)-g_{\alpha\nu}\partial_{\mu}\delta^{4}(x-y)
-\delta^{4}(x-y)\partial_{\alpha}g_{\mu\nu}\ .
\end{equation}
The gauge-space metric in Eq.~(\ref{gmetric}) is then \cite{HTC2},
\begin{equation}
\gamma_{\alpha x,\beta y}\vert_{back}=-2\delta_{\alpha\beta}
\partial^{2}\delta^{4}(x-y)\ ,
\end{equation}
and its inverse is,
\begin{equation}
\gamma^{\alpha x,\beta y}\vert_{back}=
-\frac{1}{2}\delta_{\alpha\beta}\left(\frac{1}{\partial^{2}}
\right)\delta^{4}(x-y)\ .
\end{equation}
Using these results in Eqs.~(\ref{gamma})-(\ref{qgauge}), we
have
\begin{equation}
\left. Q^{\alpha x}_{g_{\mu\nu}(y)}\right\vert_{back}=
-\frac{1}{2}\left(\frac{1}{\partial^{2}}\right)
(\delta_{\mu\alpha}\partial_{\nu}+\delta_{\nu\alpha}\partial_{\mu}
-\delta_{\mu\nu}\partial_{\alpha})\delta^{4}(x-y)\ ,
\end{equation}
and
\begin{equation}
\left. T^{g_{\rho\sigma}(z)}_{g_{\mu\nu}(x)g_{\alpha\beta}(y)}
S_{,g_{\rho\sigma}(z)}\right\vert_{back}=
-\frac{1}{2}\Lambda
(\delta_{\mu\eta}\partial_{\nu}+\delta_{\nu\eta}\partial_{\mu}
-\delta_{\mu\nu}\partial_{\eta})
\frac{1}{\partial^{2}}
(\delta_{\alpha\eta}\partial_{\beta}
+\delta_{\beta\eta}\partial_{\alpha}
-\delta_{\alpha\beta}\partial_{\eta})\delta^{4}(x-y)\ .
\label{tterm}
\end{equation}
By define
\begin{eqnarray}
{\cal L'}&\equiv&-\frac{1}{2}
h_{\mu\nu}(x)\left. (T^{g_{\rho\sigma}(z)}_{g_{\mu\nu}(x)
g_{\alpha\beta}(y)}S_{,g_{\rho\sigma}(z)})\right\vert_{back}
h_{\alpha\beta}(y)\ ,
\nonumber\\
&=&-\Lambda(h_{\mu\rho,\mu}-\frac{1}{2}h_{,\rho})
\frac{1}{\partial^{2}}(h_{\nu\rho,\nu}-\frac{1}{2}h_{,\rho})\ ,
\label{lprime}
\end{eqnarray}
the VD corrections can be accounted for by adding
${\cal L'}$ to ${\cal L}_{q}$ in Eq.~(\ref{lq}),
\begin{eqnarray}
&&{\cal L}_{VD}
\nonumber\\
&\equiv&{\cal L}_{q}+{\cal L}'
\nonumber\\
&=&\frac{1}{4}h_{\mu\nu}(-\partial^{2}-2\Lambda)
h_{\mu\nu}-\frac{1}{4}h\left[\left(1-\frac{1}{2\alpha}\right)
(-\partial^{2})-2\Lambda\right]h\nonumber\\
&&\ \ +\frac{1}{2}h_{\mu\nu}\left[
\left(\frac{1}{\alpha}-1\right)(-\partial^{2})+2\Lambda\right]
\frac{\partial_{\nu}\partial_{\rho}}{\partial^{2}}h_{\mu\rho}
-\frac{1}{2}h\left[\left(\frac{1}{\alpha}-1\right)
(-\partial^{2})+2\Lambda\right]
\frac{\partial_{\mu}\partial_{\nu}}{\partial^{2}}h_{\mu\nu}
\nonumber\\
&&\ \ \ \ +\overline{\eta}_{\mu}(-\partial^{2})\eta_{\mu}\ .
\end{eqnarray}
The corresponding ten eigenvalues for ${\cal L}_{VD}$,
as compared to Eqs.~(\ref{lambda1})-(\ref{lambda10}), are
\begin{eqnarray}
&&\lambda_{1}=\lambda_{2}=\lambda_{3}=k^{2}-2\Lambda\ ,
\\
&&\lambda_{4}=\lambda_{5}=\frac{1}{2}(k^{2}-2\Lambda)\ ,
\\
&&\lambda_{6}=\lambda_{7}=\lambda_{8}
=\left(\frac{1}{\alpha}\right)k^{2}\ ,
\\
&&\lambda_{9}=\frac{1}{2\alpha}\left\{
\left[k^{2}(1-\alpha)+2\alpha\Lambda\right]
+\sqrt{(1-\alpha+\alpha^{2})k^{4}+
2\alpha\Lambda k^{2}(1-2\alpha)+4\alpha^{2}\Lambda^{2}}
\right\}\ ,
\nonumber\\
&&
\\
&&\lambda_{10}=\frac{1}{2\alpha}\left\{
\left[k^{2}(1-\alpha)+2\alpha\Lambda\right]
-\sqrt{(1-\alpha+\alpha^{2})k^{4}+
2\alpha\Lambda k^{2}(1-2\alpha)+4\alpha^{2}\Lambda^{2}}
\right\}\ .
\nonumber\\
&&
\end{eqnarray}
Following the same procedures as in Section II,
we obtain the necessary $\zeta$-function values for the graviton
field in
the VD formalism as,
\begin{equation}
\zeta^{VD}_{gr}(0)=\frac{V_{4}}{(4\pi)^2}(12\Lambda^{2})\ ,
\end{equation}
and for the ghost field we again have
\begin{equation}
\zeta^{VD}_{gh}(0)=0\ .
\end{equation}
Therefore, the trace anomaly in the VD formalism is given
by
\begin{equation}
\langle {T_{\mu}}^{\mu}\rangle_{VD}=\frac{1}{(4\pi)^{2}}
(12\Lambda^{2})\ ,
\label{tracevd}
\end{equation}
which is independent of the gauge parameter $\alpha$. We
have thus confirmed explicitly that the trace anomaly in
the VD formalism with a flat background is gauge independent
even though this background is not a classical solution
of Einstein gravity (with cosmological constant).

Note that the usual trace anomaly in Eq.~(\ref{tracef})
will be the same as the one in the VD formalism in
Eq.~(\ref{tracevd}) if the gauge parameter $\alpha$ is set to
zero (Landau-DeWitt gauge). This is because in the
Landau-DeWitt gauge, the non-local $T_{ij}^{k}$ terms
vanish \cite{FT}. This can be easily seen in the case that we
are considering.
For the Landau-DeWitt gauge, we have basically
\begin{equation}
h_{\mu\rho}^{,\mu}-\frac{1}{2}h_{,\rho}=0\ ,
\end{equation}
and the VD correction due to the $T_{ij}^{k}$ term in
Eq.~(\ref{lprime})  clearly vanishes.
Moreover, the Christoffel symbol term
does not contribute in our case (Eq.~(\ref{christt})).
The usual trace anomaly in the Landau-DeWitt gauge
is thus identical to the VD trace anomaly.

In the next section we shall choose the Landau-DeWitt
gauge to avoid evaluating the complicated non-local $T_{ij}^{k}$
terms when evaluating trace anomalies in
more general spacetimes. Although the operators whose
$\zeta$-functions we need are simplified
in this gauge, they remain non-minimal. The
$\zeta$-functions for non-minimal operators have been
discussed in some detail in \cite{CK}.
In the next section we shall make use of those results to calculate
the trace anomaly.

\section{Vilkovisky-DeWitt trace anomalies on N-spheres}

In this section we show how to calculate the trace anomaly
in the VD formalism for a general background spacetime, and
then we consider explicitly the case of even-dimensional N-spheres.
As discussed in the last section, we shall adopt the Landau-DeWitt
gauge so that we can avoid calculating the non-local $T_{ij}^{k}$ terms.

We now return to the Lagrangian in Eq.~(\ref{action}) and consider a
general N-dimensional background
spacetime with metric $g_{\mu\nu}$. Instead of the splitting
in Eq.~(\ref{split}), we have
\begin{equation}
\bar{g}_{\mu\nu}=g_{\mu\nu}+h_{\mu\nu}\ .
\end{equation}
The quadratic part of the Lagrangian becomes
\begin{equation}
{\cal L}_{quad}=\frac{1}{2}h_{\mu\nu}
\gamma^{\mu\nu,\rho\sigma}\left[
\delta_{\rho\sigma}^{\alpha\beta}(-\Box)+
2\delta_{(\rho}^{(\alpha}\nabla_{\sigma)}\nabla^{\beta)}
-g^{\alpha\beta}\nabla_{(\rho}\nabla_{\sigma)}
-{K_{\rho\sigma}}^{\alpha\beta}\right]h_{\alpha\beta}\ ,
\label{nlquad}
\end{equation}
where
\begin{eqnarray}
\gamma^{\mu\nu,\rho\sigma}&=&\frac{1}{4}
(g^{\mu\rho}g^{\nu\sigma}+g^{\mu\sigma}g^{\nu\rho}
-g^{\mu\nu}g^{\rho\sigma})\ ,
\\
\delta_{\rho\sigma}^{\alpha\beta}&=&\delta_{\rho}^{(\alpha}
\delta_{\sigma}^{\beta)}\ ,
\\
{K_{\rho\sigma}}^{\alpha\beta}&=&
2R_{\rho\ \ \sigma}^{\ (\alpha\ \beta)}+
2\delta_{(\rho}^{(\alpha}R_{\sigma)}^{\beta)}
-g^{\alpha\beta}R_{\rho\sigma}
-\frac{2}{N-2}g_{\rho\sigma}R^{\alpha\beta}
+\frac{1}{N-2}g_{\rho\sigma}g^{\alpha\beta}R
\nonumber\\
&&\ \ -(R-2\Lambda)\delta_{\rho\sigma}^{\alpha\beta}\ .
\end{eqnarray}
The gauge-fixing part of the Lagrangian is
\begin{equation}
{\cal L}_{gf}=-\frac{1}{2\alpha}h_{\mu\nu}
\gamma^{\mu\nu,\rho\sigma}\left[
2\delta_{(\rho}^{(\alpha}\nabla_{\sigma)}\nabla^{\beta)}
-g^{\alpha\beta}\nabla_{(\rho}\nabla_{\sigma)}
\right]h_{\alpha\beta}\ ,
\label{nlgf}
\end{equation}
and the corresponding ghost Lagrangian is
\begin{equation}
{\cal L}_{gh}=\bar{\eta}_{\alpha}g^{\alpha\gamma}
\left[\delta_{\gamma}^{\beta}(-\Box)-{R_{\gamma}}^{\beta}
\right]\eta_{\beta}\ .
\label{nlgh}
\end{equation}
Combining Eqs.~(\ref{nlquad})-(\ref{nlgh}),
we obtain the quantum Lagrangian in a general
background spacetime:
\begin{eqnarray}
{\cal L}_{q}&=&\frac{1}{2}h_{\mu\nu}
\gamma^{\mu\nu,\rho\sigma}\biggl[
\delta_{\rho\sigma}^{\alpha\beta}(-\Box)+
2\left(1-\frac{1}{\alpha}\right)
\delta_{(\rho}^{(\alpha}\nabla_{\sigma)}\nabla^{\beta)}
-\left(1-\frac{1}{\alpha}\right)
g^{\alpha\beta}\nabla_{(\rho}\nabla_{\sigma)}
\nonumber\\
&&\ \ -{K_{\rho\sigma}}^{\alpha\beta}\biggr]h_{\alpha\beta}
+\bar{\eta}_{\alpha}g^{\alpha\gamma}
\left[\delta_{\gamma}^{\beta}(-\Box)-{R_{\gamma}}^{\beta}
\right]\eta_{\beta}\ .
\end{eqnarray}

To calculate the VD corrections, we need to first evaluate the
connection symbols in Eq.~(\ref{gamma}). Because we adopt
the Landau-DeWitt gauge, the non-local $T_{ij}^{k}$ terms will
not contribute, and we only have to concentrate on the
Christoffel symbols. For a general background
spacetime \cite{CK1},
\begin{eqnarray}
\left\{
\begin{array}{c}
g_{\rho\sigma}(z) \\ g_{\mu\nu}(x)g_{\alpha\beta}(y)
\end{array}
\right\}&=&\delta(x-y)\delta(y-z)\left[
\frac{1}{4}g^{\mu\nu}\delta_{\rho\sigma}^{\alpha\beta}
+\frac{1}{4}g^{\alpha\beta}\delta_{\rho\sigma}^{\mu\nu}
-\frac{1}{2}\delta_{(\rho}^{\mu}\delta_{\sigma)}^{(\alpha}
g^{\beta)\nu}\right.
\nonumber\\
&&\ \ \left.-\frac{1}{2}\delta_{(\rho}^{\nu}
\delta_{\sigma)}^{(\alpha}g^{\beta)\mu}
+\frac{1}{N-2}g_{\rho\sigma}\gamma^{\mu\nu,\alpha\beta}
\right]\ .
\end{eqnarray}
{}From Eq.~(\ref{action}),
\begin{equation}
S_{,g_{\mu\nu}(x)}=R^{\mu\nu}-\frac{1}{2}g^{\mu\nu}R+
\Lambda g^{\mu\nu}\ .
\end{equation}
The VD correction terms can be accounted for by adding
to ${\cal L}_{q}$ the
following Lagrangian,
\begin{eqnarray}
{\cal L'}&=&-\frac{1}{2}h_{\mu\nu}(x)
\left\{
\begin{array}{c}
g_{\rho\sigma}(z) \\ g_{\mu\nu}(x)g_{\alpha\beta}(y)
\end{array}
\right\}S_{,g_{\rho\sigma}(z)}h_{\alpha\beta}(y)\ ,
\nonumber\\
&=&\frac{1}{2}h_{\mu\nu}\gamma^{\mu\nu,\rho\sigma}
\left[2\delta_{(\rho}^{(\alpha}R_{\sigma)}^{\beta)}
-\frac{1}{2}g^{\alpha\beta}R_{\rho\sigma}
-\frac{1}{N-2}g_{\rho\sigma}R^{\alpha\beta}
-\frac{1}{N-2}(\delta_{\rho\sigma}^{\alpha\beta}
-\frac{1}{2}g_{\rho\sigma}g^{\alpha\beta})R\right.
\nonumber\\
&&\ \ \left.-\frac{N-4}{2(N-2)}\delta_{\rho\sigma}^{\alpha\beta}
(R-2\Lambda)\right]h_{\alpha\beta}\ .
\end{eqnarray}
The VD Lagrangian becomes
\begin{eqnarray}
{\cal L}_{VD}
&=&{\cal L}_{q}+{\cal L'}
\nonumber\\
&=&\frac{1}{2}h_{\mu\nu}
\gamma^{\mu\nu,\rho\sigma}\biggl[
\delta_{\rho\sigma}^{\alpha\beta}(-\Box)+
2\left(1-\frac{1}{\alpha}\right)
\delta_{(\rho}^{(\alpha}\nabla_{\sigma)}\nabla^{\beta)}
-\left(1-\frac{1}{\alpha}\right)
g^{\alpha\beta}\nabla_{(\rho}\nabla_{\sigma)}
\nonumber\\
&&\ \ -{P_{\rho\sigma}}^{\alpha\beta}\biggr]h_{\alpha\beta}
+\bar{\eta}_{\alpha}g^{\alpha\gamma}
\left[\delta_{\gamma}^{\beta}(-\Box)-{R_{\gamma}}^{\beta}
\right]\eta_{\beta}\ ,
\end{eqnarray}
where
\begin{eqnarray}
{P_{\rho\sigma}}^{\alpha\beta}&=&
2R_{\rho\ \ \sigma}^{\ (\alpha\ \beta)}
-\frac{1}{2}g^{\alpha\beta}R_{\rho\sigma}
-\frac{1}{N-2}g_{\rho\sigma}R^{\alpha\beta}
+\frac{1}{2(N-2)}g_{\rho\sigma}g^{\alpha\beta}R
\nonumber\\
&&\ -\frac{1}{2}\left[R-\frac{2N}{N-2}\Lambda\right]
\delta_{\rho\sigma}^{\alpha\beta}\ .
\end{eqnarray}

As in Eq.~(\ref{trace}), the trace anomaly for gravitons
in a general N-dimensional background
can now be written as:
\begin{equation}
\langle{T_{\mu}}^{\mu}\rangle_{VD}^{N}
=\left.\frac{1}{V_{N}}\left[\zeta_{M_{T}}^{N}(0)
-2\zeta_{M_{V}}^{N}(0)\right]\right\vert_{\alpha\rightarrow 0}\ ,
\label{tracegvd}
\end{equation}
where $M_{T}$ is the tensor operator for $h_{\mu\nu}$ in
${\cal L}_{VD}$,
\begin{equation}
{{M_{T}}_{\rho\sigma}}^{\alpha\beta}
=\delta_{\rho\sigma}^{\alpha\beta}(-\Box)
+2\left(1-\frac{1}{\alpha}\right)\delta_{(\rho}^{(\alpha}
\nabla_{\sigma)}\nabla^{\beta)}-\left(1-\frac{1}{\alpha}\right)
g^{\alpha\beta}\nabla_{(\rho}\nabla_{\sigma)}
-{P_{\rho\sigma}}^{\alpha\beta}\ ,
\label{tenop}
\end{equation}
and $M_{V}$ is the ghost operator,
\begin{equation}
{{M_{V}}_{\alpha}}^{\beta}=\delta_{\alpha}^{\beta}
(-\Box)-{R_{\alpha}}^{\beta}\ .
\label{vecop}
\end{equation}
Since $M_{T}$ is a non-minimal operator, it is quite difficult
to evaluate its $\zeta$-function. However, we have devised a
method in \cite{CK} to accomplish this in the case of maximally
symmetric background spacetimes. In particular, we have explicitly
given the $\zeta(0)$ values for non-minimal tensor and vector
operators on N-spheres of even dimensions 2 through 10.
In the following we use these results and evaluate
the VD trace anomalies for gravitons.

On N-spheres the Riemann tensor, the Ricci tensor, and the scalar
curvature are given by
\begin{eqnarray}
R_{\mu\nu\alpha\beta}&=&\frac{1}{r^2}(g_{\mu\alpha}g_{\nu\beta}
-g_{\mu\beta}g_{\nu\alpha})\ ,
\label{Riemann}\\
R_{\mu\nu}&=&\frac{1}{r^2}(N-1)g_{\mu\nu}\ ,
\label{RicciT}\\
R&=&\frac{1}{r^2}N(N-1)\ ,
\label{RicciS}\end{eqnarray}
where $r$ is the radius of the sphere. The operator $M_{T}$
in Eq.~(\ref{tenop}) becomes
\begin{eqnarray}
{{M_{T}}_{\rho\sigma}}^{\alpha\beta}&=&
\delta_{\rho\sigma}^{\alpha\beta}
\left[-\Box+\frac{1}{2r^{2}}(N^{2}-N+4)
-\frac{N}{N-2}\Lambda\right] -{2\over r^2}g^{\alpha\beta}g_{\rho\sigma}
\nonumber\\
&&\ +2\left(1-\frac{1}{\alpha}\right)
\delta_{(\rho}^{(\alpha}\nabla_{\sigma)}\nabla^{\beta)}
-\left(1-\frac{1}{\alpha}\right)
g^{\alpha\beta}\nabla_{(\rho}\nabla_{\sigma)}\ ,
\label{tenops}
\end{eqnarray}
and the operator $M_{V}$ in Eq.~(\ref{vecop}) becomes
\begin{equation}
{{M_{V}}_{\alpha}}^{\beta}=\delta_{\alpha}^{\beta}
\left(-\Box-\frac{N-1}{r^{2}}\right)\ .
\label{vecops}
\end{equation}

Using the results in the Appendix of \cite{CK} and taking the
Landau-DeWitt gauge ($\alpha\rightarrow 0$), the
$\zeta$-functions for $M_{T}$ on N-spheres in Eq.~(\ref{tenops})
are
\begin{eqnarray}
\zeta_{M_{T}}^{4}(0)&=&
2(\Lambda r^{2})^{2}-16(\Lambda r^{2})+\frac{299}{9}\ ,
\\
\zeta_{M_{T}}^{6}(0)&=&
\frac{9}{64}(\Lambda r^{2})^{3}-\frac{63}{16}(\Lambda r^{2})^{2}
+\frac{529}{16}(\Lambda r^{2})-\frac{2509}{36}\ ,
\\
\zeta_{M_{T}}^{8}(0)&=&
\frac{16}{3645}(\Lambda r^{2})^{4}
-\frac{1088}{3645}(\Lambda r^{2})^{3}
+\frac{584}{81}(\Lambda r^{2})^{2}
-\frac{2000}{27}(\Lambda r^{2})+\frac{45097}{150}\ ,
\\
\zeta_{M_{T}}^{10}(0)&=&
\frac{625}{8257536}(\Lambda r^{2})^{5}
-\frac{10625}{1032192}(\Lambda r^{2})^{4}
+\frac{419875}{774144}(\Lambda r^{2})^{3}
-\frac{49855}{3584}(\Lambda r^{2})^{2}
\nonumber\\
&&\ +\frac{5628229}{32256}(\Lambda r^{2})
-\frac{2242392227}{2721600}\ .
\end{eqnarray}
While the $\zeta(0)$ values of minimal operator $M_{V}$ in
Eq.~(\ref{vecops}) are,
\begin{eqnarray}
\zeta_{M_{V}}^{4}(0)&=&\frac{358}{45}\ ,
\\
\zeta_{M_{V}}^{6}(0)&=&\frac{4808}{315}\ ,
\\
\zeta_{M_{V}}^{8}(0)&=&\frac{347857}{14175}\ ,
\\
\zeta_{M_{V}}^{10}(0)&=&\frac{66840359}{1871100}\ .
\end{eqnarray}
Putting these back into Eq.~(\ref{tracegvd}), and
using the fact that the volume of a N-sphere is
\begin{equation}
V_{N}=\frac{2\pi^{(N+1)/2}}{\Gamma((N+1)/2)}\ r^{N}\ ,
\end{equation}
we obtain the VD trace anomalies for gravitons on
N-spheres:
\begin{eqnarray}
\langle{T_{\mu}}^{\mu}\rangle_{VD}^{N=4}&=&
\frac{1}{(4\pi)^{2}}\left[
12\Lambda^{2}-96\frac{\Lambda}{r^{2}}
+\frac{1558}{15}\frac{1}{r^{4}}\right]\ ,
\label{trace4}
\\
\langle{T_{\mu}}^{\mu}\rangle_{VD}^{N=6}&=&
\frac{1}{(4\pi)^{3}}\left[\frac{135}{16}\Lambda^{3}
-\frac{945}{4}\frac{\Lambda^{2}}{r^{2}}
+\frac{7935}{4}\frac{\Lambda}{r^{4}}
-\frac{42093}{7}\frac{1}{r^{6}}\right]\ ,
\label{trace6}
\\
\langle{T_{\mu}}^{\mu}\rangle_{VD}^{N=8}&=&
\frac{1}{(4\pi)^{4}}\left[\frac{896}{243}\Lambda^{4}
-\frac{60928}{243}\frac{\Lambda^{3}}{r^{2}}
+\frac{163520}{27}\frac{\Lambda^{2}}{r^{4}}
-\frac{560000}{9}\frac{\Lambda}{r^{6}}\right.
\nonumber\\
&&\ \ \left.
+\frac{5705524}{27}\frac{1}{r^{8}}\right]\ ,
\\
\langle{T_{\mu}}^{\mu}\rangle_{VD}^{N=10}&=&
\frac{1}{(4\pi)^{5}}\left[\frac{9375}{8192}\Lambda^{5}
-\frac{159375}{1024}\frac{\Lambda^{4}}{r^{2}}
+\frac{2099375}{256}\frac{\Lambda^{3}}{r^{4}}
-\frac{6730425}{32}\frac{\Lambda^{2}}{r^{6}}\right.
\nonumber\\
&&\ \ \left.+\frac{84423435}{32}\frac{\Lambda}{r^{8}}
-\frac{5361041197}{396}\frac{1}{r^{10}}\right]\ .
\label{trace10}
\end{eqnarray}
We can also obtain the trace anomalies in Euclidean
spaces by taking $r\rightarrow\infty$ in
Eqs.~(\ref{trace4})-(\ref{trace10}). Hence in Euclidean spaces,
\begin{eqnarray}
\langle{T_{\mu}}^{\mu}\rangle_{VD}^{N=4}&=&
\frac{1}{(4\pi)^{2}}(12\Lambda^{2})\ ,
\\
\langle{T_{\mu}}^{\mu}\rangle_{VD}^{N=6}&=&
\frac{1}{(4\pi)^{3}}\left(\frac{135}{16}\Lambda^{3}\right)\ ,
\\
\langle{T_{\mu}}^{\mu}\rangle_{VD}^{N=8}&=&
\frac{1}{(4\pi)^{4}}\left(\frac{896}{243}\Lambda^{4}\right)\ ,
\\
\langle{T_{\mu}}^{\mu}\rangle_{VD}^{N=10}&=&
\frac{1}{(4\pi)^{5}}\left(\frac{9375}{8192}\Lambda^{5}\right)\ .
\end{eqnarray}
Note that the $N=4$ result here agrees with the one in
Eq.~(\ref{tracevd}).

\section{Conclusions}

We have confirmed that the trace anomaly for gravitons is gauge
dependent if the background spacetime is not a solution of the
classical equations of motion. By using the VD effective action
formalism the gauge dependency was eliminated
and a unique off-shell trace anomaly
for gravitons  was obtain. Explicit evaluation of
this VD trace anomaly involved the evaluation of
$\zeta$-functions of non-minimal operators. The necessary
$\zeta(0)$ values on maximally
symmetric background
spacetimes
were given in our previous paper \cite{CK}. Using  results  obtained there we
were able to evaluate
gravitational trace anomalies on N-spheres and Euclidean
spaces (for even dimensions from 4 to 10). The 4D result of Eq. (\ref{trace4})
can be confirmed by \cite{FT}. However, the 6D result of Eq. (\ref{trace6})
does not agree with \cite{CK1}. A erratum for that paper is being prepared.

It should be straight forward to extend this calculation
to other maximally symmetric spaces, notably Kaluza-Klein spacetimes
like $M^{4}\times S^{N_{1}}\times S^{N_{2}}\times\cdots$. This
consideration is important in the discussion of the cancelation of
trace anomalies between different species of particles \cite{PTT}
in these
spacetimes. We hope to address this and related problems in
a future publication.

Although the VD effective action is manifestly gauge independent,
it possesses, as pointed out by Odintsov \cite{SDO}, an ambiguity
with respect to the choice of the field space metric. In this paper,
since we are working exclusively with Einstein gravity, we have
chosen to stay with the field metric in Eq.~(\ref{fmetric}). This
particular field metric comes out quite naturally from the Einstein
action, as derived by Vilkovisky \cite{GV}.

The method described in this paper can also be applied to the
evaluation of the Casimir energies or the one-loop effective
potentials in Kaluza-Klein spacetimes. In \cite{CK1}, we were able
to obtain the VD effective action for a general background
spacetime using a method of Barvinsky and Vilkovisky \cite{BV}.
However, due to the complexity of that calculation it seems quite
impossible to push the method to higher dimensions. On the other
hand, the procedures in this paper are much more manageable and they
can  be implemented by computer code. There should
be no major difficulty in extending them to dimensions  higher than 10.

\acknowledgements

H. T. Cho is supported by the National Science Council
of the Republic of China under contract number
NSC 84-2112-M-032-003. R. Kantowski is supported by the
U.S. Department of Energy.


\begin{references}

\bibitem[1]{GK}G. Kunstatter, in {\it Proceedings of the NATO
	       Advanced Research Workshop on Super Field Theories},
	       Burnaby, British Columbia, 1986, edited by H. C. Lee,
	       V. Elias, G. Kunstatter, R. B. Mann, and
	       K. S. Viswanathan (Plenum, New York, 1987).

\bibitem[2]{R-DS}S. Randjbar-Daemi and M. H. Sarmadi, Phys. Lett. {\bf 151B},
343 (1985).

\bibitem[3]{KL}G. Kunstatter and H. P. Leivo, Phys. Lett. B
	       {\bf 183}, 75 (1987).

\bibitem[4]{BOS}I. L. Buchbinder, S. D. Odintsov, and I. L. Shapiro, {\it
Effective Action in Quantum Gravity}, (IOP Publishing Ltd., Bristol, 1992).


\bibitem[5]{RC}R. Critchley, Phys. Rev. D{\bf 18}, 1849 (1978).


\bibitem[6]{GV}G. Vilkovisky, Nucl. Phys. {\bf B234}, 125 (1984);
	       in {\it Quantum Theory of Gravity}, edited by
	       S. M. Christensen (Hilger, Bristol, 1984),
	       pp. 169-209.
\bibitem[7]{BSD}B. S. DeWitt, in {\it Architecture of Fundamental
	       Interactions at Short Distances}, proceedings of
	       the Les Houches Summer School, Les Houches, France,
	       1985, edited by P. Ramond and R. Stora (Les Houches
	       Summer School Proceedings, Vol. 44) (North-Holland,
	       Amsterdam, 1987).

\bibitem[8]{BO}I. L. Buchbinder and S. D. Odintsov, Yad. Phys. {\bf 47}, 598
(1988); Sov. J. Nucl. Phys.{\bf 47}, 377 (1988).

\bibitem[9]{HKLT}S. R. Huggins, G. Kunstatter, H. P. Leivo, and
		 D. J. Toms, Phys. Rev. Lett. {\bf 58}, 296 (1987);
		 Nucl. Phys. {\bf B301}, 627 (1988).
\bibitem[10]{BLO}I. L. Buchbinder, P. M. Lavrov, and S. D. Odintsov,
		Nucl. Phys. {\bf B308}, 191 (1988).

\bibitem[11]{CK1}H. T. Cho and R. Kantowski, Phys. Rev. Lett. {\bf 67},
	       422 (1991).

\bibitem[12]{KM}R. Kantowski and Caren Marzban, Phys. Rev. D{\bf 46},
	       5449 (1992); E. Elizalde, S. Naftulin, and S. D. Odintsov,
		Z. Phys. C {\bf 60}, 327 (1993).

\bibitem[13]{O}S. D. Odintsov, Europhys. Lett.  {\bf 10}, 287 (1989).


\bibitem[14]{BV}A. O. Barvinsky and G. Vilkovisky, Phys. Rep. {\bf 119},
	       1 (1985).
\bibitem[15]{CK}H. T. Cho and R. Kantowski, {\it Zeta-Functions for
	       Non-Minimal Operators}, Preprint.
\bibitem[16]{HTC1}H. T. Cho, Phys. Rev. D{\bf 40}, 3302 (1989).
\bibitem[17]{LP}L. Parker, in {\it Recent Developments in Gravitation},
		edited by M. L\'evy and S. Deser (Plenum, New York, 1979).
\bibitem[18]{HTC2}H. T. Cho, Phys. Rev. D{\bf 43}, 1859 (1991).
\bibitem[19]{FT}E. S. Fradkin and A. A. Tseytlin, Nucl. Phys. {\bf B234},
	       509 (1984).
\bibitem[20]{PTT}P. Pascual, J. Taron, and R. Tarrach,
		 Phys. Rev. D{\bf 39}, 2993 (1989).
\bibitem[21]{SDO}S. D. Odintsov, Phys. Lett. B
	       {\bf 262}, 394 (1991).


\end{references}
\end{document}